\documentclass[12pt]{extarticle}
\usepackage{setspace}

\usepackage[T1]{fontenc}

\usepackage[latin1]{inputenc}
\usepackage{epsfig}
\usepackage[english]{babel}
\usepackage{color}
\usepackage{graphicx}

\usepackage{dcolumn}

\usepackage{moreverb}

\usepackage{amsmath,amssymb,amsfonts}

\begin{document}
\numberwithin{equation}{section}
\newcommand{\boxedeqn}[1]{%
  \[\fbox{%
      \addtolength{\linewidth}{-2\fboxsep}%
      \addtolength{\linewidth}{-2\fboxrule}%
      \begin{minipage}{\linewidth}%
      \begin{equation}#1\end{equation}%
      \end{minipage}%
    }\]%
}


\newsavebox{\fmbox}
\newenvironment{fmpage}[1]
     {\begin{lrbox}{\fmbox}\begin{minipage}{#1}}
     {\end{minipage}\end{lrbox}\fbox{\usebox{\fmbox}}}

\raggedbottom
\onecolumn

\parindent 8pt
\parskip 10pt
\baselineskip 16pt
\noindent\title*{{\LARGE{\textbf{Quadratic algebra approach to relativistic quantum Smorodinsky-Winternitz systems}}}}
\newline
\newline
Ian Marquette
\newline
Department of Mathematics, University of York, Heslington, York, UK. YO10 5DD
\newline
im553@york.ac.uk
\newline
\newline
There exist a relation between the Klein-Gordon and the Dirac equations with scalar and vector potentials of equal magnitude (SVPEM) and the Schrödinger equation. We obtain the relativistic energy spectrum for the four relativistic quantum Smorodinsky-Winternitz systems from their quasi-Hamiltonian and the quadratic algebras studied by Daskaloyannis in the non-relativistic context. We also apply the quadratic algebra approach directly to the initial Dirac equation for these four systems and show that the quadratic algebras obtained are the same than those obtained from the quasi-Hamiltonians. We point out how results obtained in context of quantum superintegrable systems and their polynomial algebras can be applied to the quantum relativistic case.
\section{Introduction}

In recent years, many articles were devoted to the Klein-Gordon or the Dirac equation with scalar and vector potentials of equal
magnitude (SVPEM) [1-20] (also referred in the literature as systems with spin and pseudo-spin symmetries). They are highly interesting systems with applications in nuclear physics [5]. Moreover, the Klein-Gordon or the Dirac equation with SVPEM is mathematically similar to the Schrödinger equation (i.e. the relativistic systems can be transformed into a Schrödinger-like equation called a quasi-Hamiltonian[1]). Many systems well known in context of non-relativistic quantum mechanics such the isotropic harmonic oscillator [1,4,5], the hydrogen atom [6] and one of the Smorodinsky-Winternitz systems [7] were studied. Many ring-shaped systems [8-13] were also studied. Recently, it was pointed out how the dynamical symmetries of the quasi-Hamiltonian can be used to obtain the relativistic energy spectrum [1]. 
  
The study of quantum systems allowing polynomial integrals of motion using algebraic methods began in the early days of quantum mechanics [21-23]. A systematic search for classical and quantum superintegrable systems in two-dimensional Euclidean space with two second-order integrals of motion was presented in [24] and four classes of Hamiltonians were obtained. This search was pursued on spaces of constant and nonconstant curvature [25] and also for systems with third-order integrals of motion [26-30]. Over the years many articles were devoted to superintegrability and for a detailed review we refer the reader to [31]. The study of quantum superintegrable systems by the mean of quadratic algebras and their representations was discussed by many authors [32-44]. A general quadratic [37] and cubic [28,29] algebras respectively for superintegrable systems with two second order integrals of motion and systems with a second and a third order integrals were studied. Their realizations in terms of deformed oscillator algebras and the finite dimensional unitary representations were obtained. These results were used to obtain the energy spectrum of quantum superintegrable. Systems with higher order polynomial algebras and even infinite families of Hamiltonians with polynomial algebras of arbitrary order were also studied [30,42,43]. Moreover, we presented an algebraic derivation of the energy spectrum of the generalized MICZ-Kepler system in three-dimensional Euclidean space $E_{3}$, its dual the four dimensional singular oscillator in four-dimensional Euclidean space $E_{4}$ [44] and the MICZ-Kepler system on the three sphere $S^{3}$ using the quadratic algebra approach. We thus pointed out how the quadratic algebra obtained in context of two-dimensional systems can be applied to higher dimensional systems. The quadratic algebra approach was also applied to an exactly solvable position-dependent mass Schrödinger equation in two dimenions [41].

The purpose of this paper is to study Smorodinsky-Winternitz systems in the quantum relativistic context, present an algebraic calculation of the energy spectra and also point out how results obtained in context of non relativistic quantum superintegrable systems and their polynomial algebras can also be applied to study relativistic quantum systems and used to obtain the relativistic energy spectrum. 

In Section 2, we recall a relation between the Schrödinger equation and the Dirac and Klein-Gordon equations with scalar and vector potentials of equal magnitude. In Section 3, we also recall results obtained by Daskaloyannis [37] concerning non relativistic quantum superintegrable systems with two second-order integrals of motion and their quadratic algebras. In Section 4, we consider Klein-Gordon and Dirac equations with SVPEM involving the four Smorodinsky-Winternitz systems and obtain the four corresponding quasi-Hamiltonians [1]. We present the four quadratic algebras generated by the integrals of motion of the quasi-Hamiltonians, their realizations in terms of deformed oscillator algebras, together with their finite-dimensional unitary representations and the corresponding relativistic energy spectrum. In Section 5, we use a direct approach [7] and obtain the integrals of motion of the initial Dirac equation for the four Smorodinsky-Winternitz potentials. We also obtain the four quadratic algebras generated by these integrals and show that they are equivalent to those obtained from the quasi-Hamiltonians.

\section{Klein-Gordon and Dirac equations with SPVEM}
\subsection{Klein-Gordon equation}

The two-dimensional Klein-Gordon equation with equal scalar and vector potentials (SPVEM) can be transformed into a Schrödinger-like equation which is called a quasi-Hamiltonian [1] i.e. a Schrödinger equation without any spin dependence with mass and energy depending of the relativistic energy spectrum. Let us consider the time-independent Klein-Gordon equation
\begin{equation}
(c^{2}P^{2}+(mc^{2}+V_{s}(\vec{r}))^{2}-(E - V_{v}(\vec{r}))^{2})\psi(\vec{r})=0.\label{eq1}
\end{equation}
We called $V_{s}(\vec{r})$ the scalar potential and $V_{v}(\vec{r})$ the vector potentials. When these two potentials have an equal magnitude
\begin{equation}
V_{s}(\vec{r})=V_{v}(\vec{r})=\frac{V(\vec{r})}{2},\label{eq2}
\end{equation}
the equation~\eqref{eq1} with~\eqref{eq2} becomes
\begin{equation}
((c^{2}P^{2}+(mc^{2}+E)V(\vec{r})-(E^{2}-m^{2}c^{4}))\psi(\vec{r})=0.\label{eq3}
\end{equation}
The equation~\eqref{eq3} can be written as
\begin{equation}
(\frac{P^{2}}{2\tilde{m}}+V(\vec{r}))\psi=\tilde{H}\psi=\tilde{E}\psi(\vec{r}),\label{eq4}
\end{equation}
where $\tilde{m}$ and $\tilde{E}$ are given by 
\begin{equation}
\frac{E}{c^{2}} + m=2\tilde{m},\quad E-mc^{2}=\tilde{E}, \quad E \neq -mc^{2}.\label{eq5}
\end{equation}
\subsection{Dirac equation}
The two-dimensional time-independent Dirac equation with a scalar and a vector potentials has the following form [2-14]
\begin{equation}
[c \boldsymbol{\alpha} \cdot \boldsymbol{P}+\beta (mc^{2}+V_{s}(\vec{r}))+V_{v}(\vec{r})]\psi(\vec{r})=E\psi(\vec{r}),\label{eq6}
\end{equation}
with
\begin{equation*}
\boldsymbol{P}=-i\hbar\nabla,\quad \boldsymbol{\alpha}= \begin{pmatrix} 0 & \boldsymbol{\sigma} \\ \boldsymbol{\sigma} & 0 \end{pmatrix},\quad \beta= \begin{pmatrix} I & 0 \\ 0 & -I\end{pmatrix}.
\end{equation*}
In the Pauli-Dirac representation $\psi(\vec{r})=\begin{pmatrix} \phi(\vec{r})  \\ \xi(\vec{r}) \end{pmatrix}$, we obtain the following set of coupled equations for the spinor components
\begin{equation}
c \boldsymbol{\sigma} \cdot \boldsymbol{P}\xi(\vec{r})=[E-mc^{2}-V_{s}(\vec{r})-V_{v}(\vec{r})]\phi(\vec{r}),\label{eq7}
\end{equation}
\begin{equation}
c \boldsymbol{\sigma} \cdot \boldsymbol{P}\phi(\vec{r})=[E+mc^{2}+V_{s}(\vec{r})-V_{v}(\vec{r})]\xi(\vec{r}).\label{eq8}
\end{equation}
When the scalar and the vector potentials are of equal magnitude the equation~\eqref{eq7} and~\eqref{eq8} reduce to the following system of two equations
\begin{equation}
c \boldsymbol{\sigma} \cdot \boldsymbol{P}\xi(\vec{r})=[E-mc^{2}-V(\vec{r})]\phi(\vec{r}),\quad \xi(\vec{r})=\frac{c \boldsymbol{\sigma} \cdot \boldsymbol{P}}{E+mc^{2}}\phi(\vec{r}).\label{eq9}
\end{equation}
From these two equations given by~\eqref{eq9}, we obtain that the component $\phi(\vec{r})$ satisfy the Schrödinger-like equation~\eqref{eq4} using the parameters given by equation~\eqref{eq5}. The same results are also valid for systems with scalar and vector potentials with opposite sign (pseudo-spin symmetry). It was also shown that the condition which originate the spin and pseudospin symmetries (i.e. SPVEM) in the Dirac equations are the same that produce equivalent energy spectra of relativistic spin-$\frac{1}{2}$ and spin-0 particles in the presence of vector and scalar potentials [20].

\section{Quadratic algebras}
Quadratic and more generally polynomial algebras were introduced in context of non-relativistic quantum mechanics [28-44]. Let us recall results obtained in context of quantum superintegrable systems.

Considering the case of a Hamiltonian $H$ allowing two second order integrals $A$ and $B$ (i.e. $[H,A]=[H,B]=0$), the most general quadratic algebra generated by these integrals is given by the following commutation relations [37] :
\begin{equation}
[A,B]=C,\quad \label{eq10}
\end{equation}
\[ [A,C]=\alpha A^{2} +\gamma\{A,B\}+\delta A+\epsilon B+\zeta,\]
\[ [B,C]=aA^{2}-\alpha B^{2}-\beta \{A,B\}+d A-\delta B + z.\]
The structure constants $\gamma$, $\delta$, $\beta$, $\epsilon$, $\zeta$ and $z$ of the quadratic algebra given by~\eqref{eq10} are polynomials of the Hamiltonian $H$. The Casimir operator (i.e. $[K,A]=[K,B]=[K,C]=0$) of this quadratic algebra is thus given in terms of the generators by
\begin{equation*}
K=C^{2}-\alpha\{A^{2},B\}-\gamma\{A,B^{2}\}+(\alpha \gamma -\delta)\{A,B\}+(\gamma^{2}-\epsilon)B^{2}
\end{equation*}
\[+(\gamma \delta -2 \zeta)B+\frac{2a}{3}A^{3}+(d+\frac{a\gamma}{3}+\alpha^{2})A^{2}+(\frac{a\epsilon}{3}+\alpha\delta+2z)A.\]
A deformed oscillator algebras is given by the following equations :
\begin{equation}
[N,b^{\dagger}]=b^{\dagger},\quad [N,b]=-b,\quad bb^{\dagger}=\Phi(N+1),\quad b^{\dagger}b=\Phi(N),\label{eq11}
\end{equation}
There are realizations of the quadratic algebras~\eqref{eq10} in a deformed oscillator algebra~\eqref{eq11} of the form $A=A(N)$, $B=b(N)+b^{\dagger}\rho(N)+\rho(N)b$. They are two cases.

\textbf{Case} $\gamma \neq 0$
\begin{equation*}
\rho(N)=\frac{1}{2^{12}3\gamma^{8}(N+u)(1+N+u)(1+2(N+u))^{2}},\quad A(N)=\frac{\gamma}{2}((N+u)-\frac{1}{4}-\frac{\epsilon}{\gamma^{2}},
\end{equation*}
\begin{equation*}
b(N)=-\frac{\alpha((N+u)^{2}-\frac{1}{4})}{4}+\frac{\alpha\epsilon-\delta\gamma}{2\gamma^{2}}-\frac{\alpha\epsilon^{2}-2\delta\gamma\epsilon+4\gamma^{2}\zeta}{4\gamma^{4}}\frac{1}{((N+u)^{2}-\frac{1}{4})},
\end{equation*}
\begin{equation}
\Phi(N)=-3072\gamma^{6}K(-1+2(N+u))^{2}\label{eq12}
\end{equation}
\[-48\gamma^{6}(\alpha^{2}\epsilon-\alpha\delta\gamma+a\epsilon\gamma-d\gamma^{2})(-3+2(N+u))(-1+2(N+u))^{2}(1+2(N+u))\]
\[+\gamma^{8}(3\alpha^{2}+4a\gamma(-3+2(N+u))^{2}(-1+2(N+u))^{4}(1+2(N+u))^{2}\]
\[+768(\alpha\epsilon^{2}-2\delta\epsilon\gamma+4\gamma^{2}\zeta)^{2}+32\gamma^{4}(-1+2(N+u))^{2}(-1-12(N+u)\]
\[+12(N+u)^{2})(3\alpha^{2}\epsilon^{2}-6\alpha\delta\epsilon\gamma+2a\epsilon^{2}\gamma+2\delta^{2}\gamma^{2}-4d\epsilon\gamma^{2}+8\gamma^{3}z+4\alpha\gamma^{2}\zeta)\]
\[-256\gamma^{2}(-1+2(N+u))^{2}(3\alpha^{2}\epsilon^{3}-9\alpha\delta\epsilon^{2}\gamma+a\epsilon^{3}\gamma+6\delta^{2}\epsilon\gamma^{2}-3d\epsilon^{2}\gamma^{2}\]
\[+2\delta^{2}\gamma^{4}+2d\epsilon\gamma^{4}+12\epsilon\gamma^{3}z-4\gamma^{5}z+12\alpha\epsilon\gamma^{2}\zeta-12\delta\gamma^{3}\zeta+4\alpha\gamma^{4}\zeta).\]

\textbf{Case} $\gamma=0$, $\epsilon \neq 0$
\begin{equation*}
A(N)=\sqrt{\epsilon}(N+u),\quad b(N)=-\alpha(N+u)^{2}-\frac{\delta}{\sqrt{\epsilon}}(N+u)-\frac{\zeta}{\epsilon},\quad \rho(N)=1,
\end{equation*}
\begin{equation}
\Phi(N)=\frac{1}{4}(-\frac{K}{\epsilon}-\frac{z}{\sqrt{\epsilon}}-\frac{\delta}{\sqrt{\epsilon}}\frac{\zeta}{\epsilon}+\frac{\zeta^{2}}{\epsilon^{2}})\label{eq13}
\end{equation}
\[-\frac{1}{12}(3d-a\sqrt{\epsilon}-3\alpha\frac{\delta}{\sqrt{\epsilon}}+3(\frac{\delta}{\sqrt{\epsilon}})^{2}-6\frac{z}{\sqrt{\epsilon}}+6\alpha\frac{\zeta}{\epsilon}-6\frac{\delta}{\sqrt{\epsilon}}\frac{\zeta}{\epsilon})(N+u)\]
\[+\frac{1}{4}(\alpha^{2}+d-a\sqrt{\epsilon}-3\alpha\frac{\delta}{\sqrt{\epsilon}}+(\frac{\delta}{\sqrt{\epsilon}})^{2}+2\alpha\frac{\zeta}{\epsilon})(N+u)^{2}\]
\[-\frac{1}{6}(3\alpha^{2}-a\sqrt{\epsilon}-3\alpha\frac{\delta}{\sqrt{\epsilon}})(N+u)^{3}+\frac{1}{4}\alpha^{2}(N+u)^{4}.\]

The Casimir operator $K$ can be written in terms of the Hamiltonian only. We have a energy dependent Fock space of dimension p+1 if
\begin{equation}
\Phi(p+1,u,E)=0, \quad \Phi(0,u,E)=0,\quad \phi(x)>0, \quad \forall \; x>0 \quad .\label{eq14}
\end{equation}
The Fock space is defined by
\begin{equation*}
H|\tilde{E},n>=E|E,n>,\quad N|E,n>=n|E,n> \quad b|E,0>=0.
\end{equation*}
\begin{equation*}
b^{t}|n>=\sqrt{\Phi(n+1,E)}|E,n+1>,\quad b|n>=\sqrt{\Phi(n,E)}|E,n-1>.
\end{equation*}
\subsection{Quadratic algebras and quasi-Hamiltonians}

The relation between Klein-Gordon and Dirac equations and the Schrödinger equation discussed in Section 2 allows to study relativistic systems from the quadratic algebra approach using the corresponding quasi-Hamiltonian. 

In the case of a quasi-Hamiltonian $\tilde{H}$ allowing two second order integrals $A$ and $B$ (i.e. $[\tilde{H},A]=[\tilde{H},B]=0$) the most general quadratic algebra is also given by~\eqref{eq10} and the structure constant are now polynomials of the quasi-Hamiltonian.

We have a energy dependent Fock space (depending now of $\tilde{E}$) of dimension p+1 if
\begin{equation}
\Phi(p+1,u,\tilde{E})=0, \quad \Phi(0,u,\tilde{E})=0,\quad \phi(x)>0, \quad \forall \; x>0 \quad .\label{eq15}
\end{equation}
The Fock space is now defined by
\begin{equation*}
H|\tilde{E},n>=\tilde{E}|\tilde{E},n>,\quad N|\tilde{E},n>=n|\tilde{E},n> \quad b|\tilde{E},0>=0.
\end{equation*}
\begin{equation*}
b^{t}|n>=\sqrt{\Phi(n+1,\tilde{E})}|\tilde{E},n+1>,\quad b|n>=\sqrt{\Phi(n,\tilde{E})}|\tilde{E},n-1>.
\end{equation*}
The energy spectrum $\tilde{E}$ of the quasi-Hamiltonian can be calculated from the constraints given by the equation~\eqref{eq15}. The equation~\eqref{eq5} relate the energy spectrum of the quasi-Hamiltonian and the relativistic energy spectrum ($E$). Thus, the problem of obtaining the relativistic energy spectrum of the initial Dirac or Klein-Gordon equation can done algebraically from the quadratic algebra approach. 

\section{Smorodinsky-Winternitz systems}

The four Smorodinsky-Winternitz systems [24] have the following form :

\begin{equation*}
V_{1}(x,y)=\frac{m\omega^{2}}{2}(x^{2}+y^{2})+\frac{\mu_{1}}{2mx^{2}}+\frac{\mu_{2}}{2my^{2}}
\end{equation*}
\begin{equation*}
V_{2}(x,y)=\frac{m\omega^{2}}{2}(4x^{2}+y^{2})+\frac{\mu}{2my^{2}}
\end{equation*}
\begin{equation*}
V_{3}(x,y)=\frac{k}{2m^{\frac{1}{2}}r}+\frac{1}{2m r}(\frac{\mu_{1}}{r+x}+\frac{\mu_{2}}{r-x})
\end{equation*}
\begin{equation*}
V_{4}(x,y)=\frac{k}{2m^{\frac{1}{2}}r}+\frac{\mu_{1}}{2m^{\frac{1}{4}}}\frac{\sqrt{r+x}}{r}+\frac{\mu_{2}}{2m^{\frac{1}{4}}}\frac{\sqrt{r-x}}{r}
\end{equation*}

These systems for the two-dimensional Euclidean space allows two second order integrals of motion [24]. They possess many properties quantum mechanics such the multiseparability and exact solvability [45]. Over the years many articles were devoted to these systems [45-49]. These systems were studied using the path integral approach [46] and their coherent states obtained recently [47]. Moreover, it was shown by Daskaloyannis that the integrals of motion for these four systems generate a quadratic algebra [37]. 

Let us consider the Klein-Gordon equation given by equation~\eqref{eq1} and Dirac equation given by equation~\eqref{eq6} ( with $V(\vec{r})$ as defined in~\eqref{eq2} ) with these four potentials ( i.e. $V_{1}(x,y)$, $V_{2}(x,y)$, $V_{3}(x,y)$ and $V_{4}(x,y)$ ). We obtain after a change of parameters (i.e. by replacing $\mu$, $\mu_{1}$, $\mu_{2}$, $\omega$ and $k$ by $\tilde{\mu}$, $\tilde{\mu_{1}}$, $\tilde{\mu_{2}}$, $\tilde{\omega}$ and $\tilde{k}$ as given below) the following four quasi-Hamiltonians which have the same form than the non-relativistic systems with parameters depending of the relativistic energy spectrum :

\textbf{Case 1}:

\begin{equation*}
\tilde{H_{r}}=\frac{P_{x}^{2}}{2\tilde{m}}+\frac{P_{y}^{2}}{2\tilde{m}}+\frac{\tilde{m}\tilde{\omega}^{2}}{2}(x^{2}+y^{2})+\frac{\tilde{\mu_{1}}}{2\tilde{m}x^{2}}+\frac{\tilde{\mu_{2}}}{2\tilde{m}y^{2}},\quad \tilde{\omega}=\sqrt{\frac{m}{\tilde{m}}}\omega,\quad \tilde{\mu_{i}}=\frac{\tilde{m}}{m}\mu_{i},
\end{equation*}
\begin{equation*}
A_{r}=P_{x}^{2}+\tilde{m}^{2}\tilde{\omega}^{2}x^{2}+\frac{\tilde{\mu}_{1}}{x^{2}},\quad B_{r}=L^{2}+r^{2}(\frac{\tilde{\mu}_{1}}{x^{2}}+\frac{\tilde{\mu}_{2}}{y^{2}}),\quad L=xP_{y}-yP_{x}.
\end{equation*}

\textbf{Case 2}:

\begin{equation*}
\tilde{H_{r}}=\frac{P_{x}^{2}}{2\tilde{m}}+\frac{P_{y}^{2}}{2\tilde{m}}+\frac{\tilde{m}\tilde{\omega}^{2}}{2}(4x^{2}+y^{2})+\frac{\tilde{\mu}}{2\tilde{m}y^{2}},\quad \tilde{\omega}=\sqrt{\frac{m}{\tilde{m}}}\omega,\quad \tilde{\mu}=\frac{\tilde{m}}{m}\mu,
\end{equation*}
\begin{equation*}
A_{r}=P_{x}^{2}+4\tilde{m}^{2}\tilde{\omega}^{2}x^{2},\quad B_{r}=\frac{1}{2}\{L,P_{y}\}+\frac{\tilde{\mu}x}{y^{2}}-\tilde{\omega}^{2}\tilde{m}^{2}xy^{2}.
\end{equation*}

\textbf{Case 3}:

\begin{equation*}
\tilde{H_{r}}=\frac{P_{x}^{2}}{2\tilde{m}}+\frac{P_{y}^{2}}{2\tilde{m}}+\frac{\tilde{k}}{2\tilde{m}^{\frac{1}{2}}r}+\frac{1}{2\tilde{m} r}(\frac{\tilde{\mu}_{1}}{r+x}+\frac{\tilde{\mu}_{2}}{r-x}),\quad \tilde{k}=\sqrt{\frac{\tilde{m}}{m}}k,\quad \tilde{\mu_{i}}=\frac{\tilde{m}}{m}\mu_{i},
\end{equation*}
\begin{equation*}
A_{r}=L^{2}+r(\frac{\tilde{\mu}_{1}}{r+x}+\frac{\tilde{\mu}_{2}}{r-x})=\frac{1}{2}(\frac{1}{2}(\eta P_{\xi}-\xi P_{\eta})^2+(\xi^{2}+\eta^{2})(\frac{\mu_{1}}{\xi^{2}}+\frac{\mu_{2}}{\eta^{2}})),
\end{equation*}
\begin{equation*} B_{r}=\frac{1}{2}(\{L,P_{y}\}-\frac{\tilde{\mu}_{1}}{r}\frac{r-x}{r+x}+\frac{\tilde{\mu}_{2}}{r}\frac{r+x}{r-x}+\frac{\tilde{k}m^{\frac{1}{2}}x}{r})
\end{equation*}
\[=\frac{1}{\xi^{2}+\eta^{2}}(\frac{1}{2}(\xi^{2}P_{\eta}^{2}-\eta^{2}P_{\xi}^{2})+\frac{\mu_{2}\xi^{2}}{\eta^{2}}-\frac{\mu_{1}\eta^{2}}{\xi^{2}}+\frac{k}{2}(\xi^{2}-\eta^{2})). \]
with $x=\frac{1}{2}(\xi^{2}-\eta^{2})$ and $y=\xi \eta$.

\textbf{Case 4}:

\begin{equation*}
\tilde{H_{r}}=\frac{P_{x}^{2}}{2\tilde{m}}+\frac{P_{y}^{2}}{2\tilde{m}}+\frac{\tilde{k}}{2\tilde{m}^{\frac{1}{2}}r}+\frac{\tilde{\mu}_{1}}{2\tilde{m}^{\frac{1}{4}}}\frac{\sqrt{r+x}}{r}+\frac{\tilde{\mu}_{2}}{2\tilde{m}^{\frac{1}{4}}}\frac{\sqrt{r-x}}{r},\quad \tilde{k}=\sqrt{\frac{\tilde{m}}{m}}k,\quad \tilde{\mu_{i}}=\frac{\tilde{m}^{\frac{1}{4}}}{m^{\frac{1}{4}}}\mu_{i},
\end{equation*}
\begin{equation*}
A_{r}=\frac{1}{2}(-\{L,P_{y}\}+\frac{\tilde{\mu}_{1}(r-x)\sqrt{r+x}}{r}\tilde{m}^{\frac{3}{4}}-\frac{\tilde{\mu}_{2}(r+x)\sqrt{r-x}}{r}\tilde{m}^{\frac{3}{4}}-\frac{\tilde{k}\tilde{m}^{\frac{1}{2}}x}{r}),
\end{equation*}
\[=\frac{1}{2(\xi^{2}+\eta^{2})}(\eta^{2}P_{\xi}^{2}-\xi^{2}P_{\eta}^{2}+k(\eta^{2}-\xi^{2})+2\xi\eta(\mu_{1}\eta -\mu_{2}\xi)),\]
\begin{equation*}
B_{r}=\frac{1}{2}(\{L,P_{x}\}-\frac{\tilde{\mu}_{1}x\sqrt{r-x}}{r}\tilde{m}^{\frac{3}{4}}+\frac{\tilde{\mu}_{2}x\sqrt{r+x}}{r}\tilde{m}^{\frac{3}{4}}-\frac{\tilde{k}\tilde{m}^{\frac{1}{2}}y}{r}),
\end{equation*}
\[=-\frac{1}{2(\xi^{2}+\eta^{2})}(\xi\eta(P_{\xi}^{2}+P_{\eta}^{2})-(\xi^{2}+\eta^{2})P_{\xi}P_{\eta}+2k\xi\eta+(\mu_{2}\xi-\mu_{1}\eta)(\eta^{2}-\xi^{2})).\] 
 
\subsection{Quadratic algebras and relativistic energy spectum}
As discussed in the Section 3, we can use the quadratic algebra approach [37] and the realizations of quadratic algebras as deformed oscillator algebras introduced in context of non relativistic quantum mechanics to obtain the relativistic energy spectrum of the Dirac or Klein-Gordon equations with SVPEM. 

Let us present these quadratic algebras generated by the integrals of motion of the four quasi-Hamiltonians. 

\textbf{Case 1} :

\begin{equation}
[A_{r},B_{r}]=C_{r},\label{eq16} 
\end{equation}
\[[A_{r},C_{r}]=8\hbar^{2}A_{r}^{2}-16\hbar^{2}\tilde{m}\tilde{H_{r}}A_{r}+16\hbar^{2}\tilde{m}^{2}\tilde{\omega}^{2}B_{r}
-16\hbar^{2}(\tilde{\mu}_{1}+\tilde{\mu}_{2})\tilde{m}^{2}\tilde{\omega}^{2}+8\hbar^{4}\tilde{m}^{2}\tilde{\omega}^{2},\] \[[B_{r},C_{r}]=-8\hbar^{2}\{A_{r},B_{r}\}+16\hbar^{4}A_{r}+16\hbar^{2}\tilde{m}\tilde{H_{r}}B_{r}-16\hbar^{2}(\tilde{\mu}_{2}-\tilde{\mu}_{1})\tilde{m}\tilde{H_{r}}-16\hbar^{4}\tilde{m}\tilde{H_{r}}.\]
\begin{equation*}
K_{r}=16\hbar^{2}((\tilde{\mu}_{2}-\tilde{\mu}_{1})^{2}\tilde{m}^{2}\tilde{\omega}^{2}+4\tilde{\mu}_{1}\tilde{m}^{2}\tilde{H_{r}}^{2})-16\hbar^{4}( 3\tilde{m}^{2}\tilde{H_{r}}^{2}+2\hbar^{2}\tilde{m}^{2}\tilde{\omega}^{2}-2(\tilde{\mu}_{1}+\tilde{\mu}_{2})).
\end{equation*}
We introduce $\tilde{\mu}_{1}=(\tilde{k}_{1}^{2}-\frac{1}{4})\hbar^{2}$ and $\tilde{\mu}_{2}=(\tilde{k}_{2}^{2} -\frac{1}{4})\hbar^{2}$. With the constraints given by the equation~\eqref{eq15} (with $u=\frac{1}{2}+\frac{\epsilon_{1}\tilde{k}_{1}}{2}$) we obtain the energy spectrum $\tilde{E_{r}}$ of the quasi-Hamiltonian $\tilde{H_{r}}$
\begin{equation}
\tilde{E_{r}}=2\hbar\tilde{\omega}(p+1+\frac{\epsilon_{1}\tilde{k}_{1}+\epsilon_{2}\tilde{k}_{2}}{2}),\quad 
\Phi=16\hbar^{4}x(p+1-x)(x+\epsilon_{1}\tilde{k}_{1})(x+\epsilon_{1}\tilde{k}_{2}).\label{eq17}
\end{equation}

Thereby, from the equation~\eqref{eq5} and~\eqref{eq17} we can thus obtain the corresponding quantum relativistic energy spectrum $E_{r}$ for the Klein-Gordon and Dirac equal with scalar and vector potential of equal magnitude
\begin{equation}
(E_{r}-mc^{2})^{2}(\frac{E_{r}}{2c^{2}}+\frac{m}{2})=4\hbar^{2}m^{2}\omega^{2}(p+1+\epsilon_{1}\sqrt{\frac{(\frac{E_{r}}{2c^{2}}+\frac{m}{2})\mu_{1}}{m\hbar^{2}}+\frac{1}{4}}+\epsilon_{2}\sqrt{\frac{(\frac{E_{r}}{2c^{2}}+\frac{m}{2})\mu_{2}}{m\hbar^{2}}+\frac{1}{4}})^{2}.\label{eq18}
\end{equation}
As previously observed before for such systems [1-20] the relativistic energy spectrum is not given explicitly. These results coincide with the relativistic energy spectrum obtained in Ref 7 for this system (but not in its full generality i.e. with $m=c=\hbar=1$ and $\mu_{2}=0$ ).

\textbf{Case 2}:

\begin{equation}
[A_{r},B_{r}]=C_{r},\quad [A_{r},C_{r}]=16\hbar^{2}\tilde{m}^{2}\tilde{\omega}^{2}B_{r},\label{eq19}
\end{equation}
\begin{equation*}
[B_{r},C_{r}]=6\hbar^{2}A_{r}^{2}-16\hbar^{2}\tilde{m}\tilde{H_{r}}A_{r}-8\hbar^{2}(\tilde{\mu}\tilde{\omega}^{2}-\tilde{H_{r}}^{2})\tilde{m}^{2}+6\hbar^{4}\tilde{m}^{2}\tilde{\omega}^{2},
\end{equation*}
\begin{equation*}
K_{r}=64\hbar^{4}\tilde{m}^{3}\tilde{\omega}^{2}\tilde{H_{r}}.
\end{equation*}
We introduce $\tilde{\mu}=(\tilde{k}^{2}-\frac{1}{4})\hbar^{2}$. The structure function and the energy and structure function are given by (with $u=\frac{1}{2}$)
\begin{equation}
\tilde{E_{r}}=2\hbar\tilde{\omega}(p+1+\frac{\epsilon \tilde{k}}{2}\hbar),\quad \Phi=4\hbar^{3}x(p+1-x)(p+1-x+\epsilon \tilde{k})\label{eq20}
\end{equation}
From the~\eqref{eq5} and~\eqref{eq20} we obtain the relativistic energy spectrum $E_{r}$
\begin{equation}
(E_{r}-mc^{2})^{2}(\frac{E_{r}}{2c^{2}}+\frac{m}{2})=4\hbar^{2}m^{2}\omega^{2}(p+1+\frac{\epsilon}{2}\sqrt{\frac{(\frac{E_{r}}{2c^{2}}+\frac{m}{2})\mu}{m\hbar^{2}}+\frac{1}{4}})^{2}.\label{eq21}
\end{equation}

\textbf{Case 3}:

\begin{equation}
[A_{r},B_{r}]=C_{r},\label{eq22}
\end{equation}
\begin{equation*} [A_{r},C_{r}]=2\hbar^{2}\{A_{r},B_{r}\}-\hbar^{4}B_{r}-\hbar^{2}\tilde{k}\tilde{m}^{\frac{1}{2}}(\tilde{\mu}_{1}-\tilde{\mu}_{2}),
\end{equation*}
\begin{equation*}
[B_{r},C_{r}]=-2\hbar^{2}B_{r}^{2}+8\hbar^{2}\tilde{m}\tilde{H_{r}}A_{r}+\hbar^{4}\tilde{m}H_{r}-\hbar^{2}(4(\tilde{\mu}_{1}+\tilde{\mu}_{2})\tilde{H_{r}}\tilde{m}-\frac{\tilde{k}^{2}\tilde{m}}{2}),
\end{equation*}
\begin{equation*}
K_{r}=-\hbar^{2}(2 (\tilde{\mu}_{1}-\tilde{\mu}_{2})^{2}\tilde{m}\tilde{H_{r}}-\tilde{k}^{2}\tilde{m}(\tilde{\mu}_{1}+\tilde{\mu}_{2}))-2\hbar^{4}( (\tilde{\mu}_{1}+\tilde{\mu}_{2})\tilde{m} \tilde{H_{r}}-\frac{\tilde{k}^{2}\tilde{m}}{2})+\hbar^{6}\tilde{m}\tilde{H_{r}}.
\end{equation*}
We introduce $\tilde{\mu}_{1}=\frac{\hbar^{2}}{2}(\tilde{k}_{1}^{2}-\frac{1}{4})$ and $\tilde{\mu}_{2}=\frac{\hbar^{2}}{2}(\tilde{k}_{2}^{2}-\frac{1}{4})$. The structure function is given by (with $u=\frac{1}{2}(2+\epsilon_{1}\tilde{k}_{1}+\epsilon_{2}\tilde{k}_{2})$)
\begin{equation}
\tilde{E_{r}}=\frac{-\tilde{k}^{2}}{2\hbar^{2}(2(p+1)+\epsilon_{1}\tilde{k}_{1}+\epsilon_{2}\tilde{k}_{2})^{2}},\quad \Phi=2^{20}3\tilde{k}^{2}\tilde{m}^{2}\hbar^{16}x(p+1-x)(x+\epsilon_{1}\tilde{k_{1}})(x+\epsilon_{2}\tilde{k}_{2})\label{eq23}
\end{equation}
\[ (x+\epsilon_{1}\tilde{k}_{1}+\epsilon_{2}\tilde{k}_{2})\frac{(x+p+1+\epsilon_{1}\tilde{k}_{1}+\epsilon_{2}\tilde{k}_{2})}{(2(p+1)+\epsilon_{1}\tilde{k}_{1}+\epsilon_{2}\tilde{k}_{2})^{2}}.\]
From the equation~\eqref{eq5} and~\eqref{eq23} we obtain the relativistic energy spectrum $E_{r}$
\begin{equation}
(E_{r}-mc^{2})2m\hbar^{2}(2(p+1)+\epsilon_{1}\sqrt{\frac{(\frac{E_{r}}{2c^{2}}+\frac{m}{2})\mu_{1}}{m\hbar^{2}}+\frac{1}{4}}+\epsilon_{2}\sqrt{\frac{(\frac{E_{r}}{2c^{2}}+\frac{m}{2})\mu_{2}}{m\hbar^{2}}+\frac{1}{4}})^2=-(\frac{E_{r}}{2c^{2}}+\frac{m}{2})k.\label{eq24}
\end{equation}

\textbf{Case 4}:

\begin{equation}
[A_{r},B_{r}]=C_{r},\quad [A_{r},C_{r}]=-2\hbar^{2}\tilde{m}\tilde{H_{r}}B_{r}-\hbar^{2}\frac{\tilde{\mu}_{1}\tilde{\mu}_{2}}{2}\tilde{m},\label{eq25}
\end{equation}
\begin{equation*}
[B_{r},C_{r}]=2\hbar^{2}\tilde{m}\tilde{H_{r}}A_{r}-\frac{\hbar^{2}(\tilde{\mu}_{1}^{2}-\tilde{\mu}_{2}^{2})}{4}\tilde{m}
\end{equation*}
\begin{equation*}
K_{r}=\hbar^{2}\frac{\tilde{m}^{2}}{2}\tilde{k}^{2}\tilde{H_{r}}+\hbar^{2}\tilde{m}^{2}\tilde{k}\frac{(\tilde{\mu}_{1}^{2}+\tilde{\mu}_{2}^{2})}{4}+\hbar^{4}\tilde{m}^{2}\tilde{H_{r}}^{2}.
\end{equation*}

The structure function is given by (with $u=\epsilon \frac{\tilde{\mu}_{i}^{2}}{2\hbar(\sqrt{-2\tilde{E_{r}}})^{3}}
-\frac{\tilde{k}}{2\hbar(\sqrt{-2\tilde{E_{r}}})}+\frac{\epsilon}{2}$)
\begin{equation}
2\hbar(p+1)(-2\tilde{E_{r}})^{\frac{3}{2}}+4\epsilon \tilde{m}\tilde{k}\tilde{E_{r}}^{2}+\epsilon (\tilde{m})^{\frac{3}{2}}(\tilde{\mu_{1}}^{2}+\tilde{\mu_{2}}^{2})=0,\quad \Phi=-\frac{\tilde{E_{r}}\tilde{m}}{2}x(p+1-x).\label{eq26}
\end{equation}
From the equation~\eqref{eq5} and~\eqref{eq26} we obtain the relativistic energy spectrum $E_{r}$
\begin{equation}
2^{\frac{5}{2}}m^{\frac{1}{2}}(p+1)\hbar (-(E_{r}-mc^{2})(\frac{E_{r}}{2c^{2}} + \frac{m}{2}))^{\frac{3}{2}}+4\epsilon k (\frac{E_{r}}{2c^{2}} + \frac{m}{2})^{2}(E_{r}-mc^{2})+\epsilon (\frac{E_{r}}{2c^{2}} + \frac{m}{2})^{2}(\mu_{1}^{2}+\mu_{2}^{2})=0.\label{eq27}
\end{equation}

\section{Symmetry algebras of Dirac equations} 

In Section 4, we obtained the relativistic energy spectrum of the Klein-Gordon and Dirac equations with SVPEM for the four Smorodinsky-Winternitz systems using the corresponding quasi-Hamiltonians and the quadratic algebras approach. In this Section, we will obtain the symmetry algebra directly of the Dirac equation given by equation~\eqref{eq6} for these four systems and show that these quadratic algebras are the same than those obtained for the quasi-Hamiltonians. 

Let us consider as in Ref.7 integrals of the following form

\begin{equation}
Q_{d}= \begin{pmatrix} Q_{11} & Q_{12}M  \\ M^{\dagger}Q_{21} & M^{\dagger}Q_{22}M  \end{pmatrix},\label{eq28}
\end{equation}
\begin{equation}
M=P_{x}-iP_{y}.\label{eq29}
\end{equation}
The condition $[Q_{d},H]=0$ give the following constraints
\begin{equation}
Q_{12}=Q_{21},\quad Q_{11}=\frac{1}{c}Q_{12}(2+V)+Q_{11}M^{2},\label{eq30}
\end{equation}
\begin{equation}
[Q_{11},V]+c[Q_{12},M^{2}]=0,\quad [Q_{12},V]+c[Q_{22},M^{2}]=0.\label{eq31}
\end{equation}

By a direct calculation, we obtain that the initial Dirac equation for the four Smorodinsky-winternitz systems allow two integrals of motion $A_{d}$ and $B_{d}$ of the form given by the equation~\eqref{eq28} i.e.
\begin{equation}
A_{d}= \begin{pmatrix} A_{11} & A_{12}M  \\ M^{\dagger}A_{21} & M^{\dagger}A_{22}M  \end{pmatrix},\label{eq32}
\end{equation}
\begin{equation}
B_{d}= \begin{pmatrix} B_{11} & B_{12}M  \\ M^{\dagger}B_{21} & M^{\dagger}B_{22}M  \end{pmatrix},\label{eq33}
\end{equation}

where the components of $A_{d}$ and $B_{d}$ are closely related to the integrals of motion of the quasi-Hamiltonians and thus of the quantum non relativistic systems.

\textbf{Case 1}

The components of the two integrals $A_{d}$ and $B_{d}$ given by equation~\eqref{eq32} and~\eqref{eq33} are given by the following equations

\begin{equation}
A_{12}=m^{2}\omega^{2}x^{2}+\frac{\mu_{1}}{x^{2}},\quad A_{22}=\frac{2mc}{M^{2}}P_{x}^{2},
\end{equation}
\begin{equation}
B_{12}(x^{2}+y^{2})(\frac{\mu_{1}}{x^{2}}+\frac{\mu_{2}}{y^{2}}),\quad B_{22}=\frac{2cm}{M^{2}}(xP_{y}-yP_{x})^{2}.
\end{equation}

We obtain the following quadratic algebra:

\begin{equation}
[A_{d},B_{d}]=C_{d},\quad 
\end{equation}
\begin{equation}
[A_{d},C_{d}]=16c\hbar^{2}mA_{d}^{2}-32\hbar^{2}m^{2}H^{2}A_{d}+32c^{4}\hbar^{2}m^{4}A_{d}-\frac{32\hbar^{2}\omega^{2}m^{3}(\mu_{1}+\mu_{2})}{c}H^{2}+32\hbar^{2}m^{3}\omega^{2}HB_{d}
\end{equation}
\[+32c^{2}\hbar^{2}m^{4}\omega^{2}B_{d}+32c\hbar^{2}m^{4}(\hbar^{2}-2\mu_{1}-2\mu_{2})\omega^{2}H-32c^{3}\hbar^{2}m^{5}(-\hbar^{2}+\mu_{1}+\mu_{2})\omega^{2},\]
\begin{equation}
[B_{d},C_{d}]=-16c\hbar^{2}m\{A_{d},B_{d}\}+32\hbar^{2}m^{2}H^{2}B_{d}+64c^{2}m^{2}\hbar^{4}A_{d}-32c^{4}\hbar^{2}m^{4}B_{d}+\frac{32\hbar^{2}m^{2}(\mu_{1}-\mu_{2})}{c}H^{3}
\end{equation}
\[-32c\hbar^{2}m^{3}(2\hbar^{2}-\mu_{1}+\mu_{2})H^{2}-32c^{3}\hbar^{2}m^{4}(\mu_{1}-\mu_{2})H+32c^{5}\hbar^{2}m^{5}(2\hbar^{2}+\mu_{2}-\mu_{1}).\]

\textbf{Case 2}

The components of the two integrals are given by

\begin{equation}
A_{12}=m^{2}\omega^{2}4x^{2},\quad A_{22}=\frac{2mc}{M^{2}}P_{x}^{2},
\end{equation}
\begin{equation}
B_{12}=\frac{\mu x}{y^{2}}-\omega^{2}m^{2}xy^{2}, \quad B_{22}=\frac{mc}{M^{2}}\{P_{y},L\}.
\end{equation}

We obtain the quadratic algebra:

\begin{equation}
[A_{d},B_{d}]=C_{d},\quad [A_{d},C_{d}]=32\hbar^{2}m^{3}\omega^{2}HB_{d}+32c^{2}\hbar^{2}m^{4}\omega^{2}B_{d},
\end{equation}
\begin{equation}
[B_{d},C_{d}]=12c\hbar^{2}mA_{d}^{2}-32\hbar^{2}m^{2}H^{2}A_{d}+16\frac{\hbar^{2}m^{3}}{c}H^{4}+32c^{4}\hbar^{2}m^{4}A_{d}
\end{equation}
\[-\frac{16(2c^{4}\hbar^{2}m^{5}+\hbar^{2}m^{3}\mu\omega^{2})}{c}H^{2} +8c\hbar^{2}m^{4}\omega^{2}(3\hbar^{2}-4\mu)H+8c^{3}\hbar^{2}m^{3}(2c^{4}m^{4} +3\hbar^{2}m^{2}\omega^{2}-2m^{2}\mu\omega^{2}).\]

\textbf{Case 3}

The components of the integrals of motion are given by the following equations 
:
\begin{equation}
A_{12}=\frac{\xi^{2}+\eta^{2}}{2}(\frac{\mu_{1}}{\xi^{2}}+\frac{\mu_{2}}{\eta^{2}}),\quad A_{22}=-\frac{c\hbar^{2}m}{2M^{2}}(\eta \partial_{\xi}-\xi^{2}\partial_{\eta})^{2},
\end{equation}
\begin{equation}
B_{12}=\frac{1}{\xi^{2}+\eta^{2}}(m^{\frac{1}{2}}\frac{k}{2}(\xi^{2}-\eta^{2})-\frac{\mu_{1}\eta^{2}}{\xi^{2}}+\frac{\mu_{2}\xi^{2}}{\eta^{2}}),\quad B_{22}=-\frac{2cm\hbar^{2}}{M^{2}}\frac{1}{2(\xi^{2}+\eta^{2})}(\xi^{2}\partial_{\eta}^{2}-\eta^{2}\partial_{\xi}^{2}).
\end{equation}

We obtain the quadratic algebra:

\begin{equation}
[A_{d},B_{d}]=C_{d},\quad [A_{d},C_{d}]=4c\hbar^{2}m\{A_{d},B_{d}\}-4c^{2}m^{2}\hbar^{4}B_{d}-\frac{2\hbar^{2}km^{\frac{3}{2}}(\mu_{1}-\mu_{2})}{c}H^{2}
\end{equation}
\[+4c\hbar^{2}km^{\frac{5}{2}}(\mu_{2}-\mu_{1})H-2c^{3}\hbar^{2}km^{\frac{7}{2}}(\mu_{1}-\mu_{2}),\]
\begin{equation}
[B_{d},C_{d}]=-4cm\hbar^{2}B_{d}^{2}+16\hbar^{2}m^{2}H^{2}A_{d}-\frac{8\hbar^{2}m^{2}(\mu_{1}+\mu_{2})}{c}H^{3}+\frac{\hbar^{2}m^{2}(k^{2}+4c^{2}m(\hbar^{2}-2(\mu_{1}+\mu_{2})))}{c}H^{2}
\end{equation}
\[+2c\hbar^{2}m^{3}(k^{2}+4c^{2}m(\mu_{1}+\mu_{2}))H-16c^{4}\hbar^{2}m^{4}A_{d}+c^{3}\hbar^{2}m^{4}(k^{2}+mc^{2}(-4\hbar^{2}+8(\mu_{1}+\mu_{2}))).\]

\textbf{Case 4}

We have for this system the following components for the integrals of motion:

\begin{equation}
A_{12}=\frac{1}{2(\xi^{2}+\eta^{2})}(km^{\frac{1}{2}}(\eta^{2}-\xi^{2})+2m^{\frac{1}{4}}\xi\eta (\mu_{1}\eta-\mu_{2}\xi)),
\end{equation}
\begin{equation}
A_{22}=\frac{-cm\hbar^{2}}{M^{2}}\frac{1}{\xi^{2}+\eta^{2}}(\eta^{2}\partial_{\xi}^{2}-\xi^{2}\partial_{\eta}^{2}),
\end{equation}
\begin{equation}
B_{12}=\frac{1}{2(\xi^{2}+\eta^{2})}(2km^{\frac{1}{2}}\xi\eta+m^{\frac{1}{4}}(\mu_{2}\xi-\mu_{1}\eta)(\eta^{2}-\xi^{2})),
\end{equation}
\begin{equation}
B_{22}=\frac{-cm\hbar^{2}}{M^{2}}(\frac{1}{\xi^{2}+\eta^{2}})(\xi\eta(\partial_{\xi}^{2}+\partial_{\eta}^{2})-(\xi^{2}+\eta^{2})\partial_{\xi}\partial_{\eta}).
\end{equation}

We obtain the quadratic algebra:

\begin{equation}
[A_{d},B_{d}]=C_{d},\quad [A_{d},C_{d}]=-4\hbar^{2}m^{2}H^{2}B_{d}+4c^{4}\hbar^{2}m^{4}B_{d}-\frac{\hbar^{2}m^{\frac{5}{2}}\mu_{1}\mu_{2}}{c}H^{2}-2c\hbar^{2}m^{\frac{7}{2}}\mu_{1}\mu_{2}H-c^{3}\hbar^{2}m^{\frac{9}{2}}\mu_{1}\mu_{2},
\end{equation}
\begin{equation}
[B_{d},C_{d}]=4\hbar^{2}m^{2}H^{2}A_{d}-4c^{4}\hbar^{2}m^{4}A_{d}-\frac{\hbar^{2}m^{\frac{5}{2}}(\mu_{1}^{2}-\mu_{2}^{2}}{2c}H^{2}+c\hbar^{2}m^{\frac{7}{2}}(\mu_{2}^{2}-\mu_{1}^{2})H-\frac{1}{2}c^{3}\hbar^{2}m^{\frac{9}{2}}(\mu_{1}^{2}-\mu_{2}^{2}).
\end{equation}

Considering the equation~\eqref{eq5} and the following relations $A_{d}=2mcA_{r}$, $B_{d}=2mcB_{r}$ and $C_{d}=4m^{2}c^{2}C_{r}$ we obtain that these four quadratic algebras are in fact equivalent to the one obtained for the quasi-Hamiltonians. 

\section{Conclusion}

The main result is the application of the quadratic algebra approach in the quantum relativistic context. We studied Klein-Gordon and Dirac with scalar and vector potentials of equal magnitude involving the four Smorodinsky-Winternitz systems using the quadratic algebras of the corresponding quasi-Hamiltonians. We thus obtained the relativistic energy spectrum for these systems. We also applied directly the quadratic algebra approach to the initial Dirac equation for these four potentials. We also showed that quadratic algebras generated by the integrals of the quasi-Hamiltonians and the integrals of initial Dirac equations coincide.

We pointed out also how the study of quantum superintegrable systems and their polynomial algebras is also interesting in regard of relativistic quantum mechanics. 
 
Supersymmetric quantum mechanics for the Dirac and Klein-Gordon equation was discussed in many articles [14-20,50]. For quantum superintegrable systems a relation between integrals of motion, ladder operators and supersymmetric quantum mechanics was pointed out [28-31]. It would be interesting to obtain a such relation in the quantum relativistic context.

\textbf{Acknowledgments} We thank Niall MacKay for discussions and a careful reading of the manuscript. The research of I.M. was supported by a postdoctoral research fellowship from FQRNT of Quebec.


\end{document}